\documentclass[twocolumn,showpacs,preprintnumbers,amsmath,amssymb]{revtex4}
\usepackage{dcolumn}
\usepackage{bm}
\usepackage{graphicx}
\usepackage{amsmath}
\begin{document}

\title{A novel exact solution to transmission problem of electron wave in a nonlinear Kronig-Penney superlattice}
\author{Chao Kong,\ Kuo Hai,\ Jintao Tan,\ Hao Chen,\ Wenhua Hai\footnote{whhai2005@aliyun.com}}
\affiliation{Department of Physics and Key Laboratory of
Low-dimensional Quantum Structures and\\ Quantum Control of Ministry
of Education, Hunan Normal University, Changsha 410081, China}

\begin{abstract}
Nonlinear Kronig-Penney model has been frequently employed to study transmission problem of electron wave in a nonlinear electrified chain or in a doped semiconductor superlattice. Here from an integral
equation we derive a novel exact solution of the problem, which contains a simple
nonlinear map connecting transmission coefficient with system parameters. Consequently, we suggest
a scheme for manipulating electronic distribution and transmission
by adjusting the system parameters. A new effect of quantum coherence is evidenced in the strict expression of transmission coefficient by which for some
different system parameters we obtain the similar aperiodic
distributions and arbitrary transmission coefficients including the
approximate zero transmission and total transmission, and the multiple transmissions. The method
based on the concise exact solution can be applied directly to some nonlinear cold atomic
systems and a lot of linear Kronig-Penney systems, and also can be extended to investigate electron transport in different
discrete nonlinear systems.


\end{abstract}

\maketitle

\section{Introduction}

With the advances of new technology and experimental techniques in
material science, more and more artificial materials with special
structures and properties have been invented, such as a variety of semiconductor
heterostructures and strained layer superlattice
materials \cite{Hennig,Alferov,Lamberti}. As we known, lattice
constants of many superlattice materials are in the same order of magnitude
as electron wavelength, so quantum effect becomes significant in the related
works. Research on electron wave
propagation through a series of potential barriers, as a
fundamental problem of quantum mechanics, is widely
applied to study electronic transport properties
in synthetic materials, including the electron wave
propagation in doped semiconductor superlattice materials with nonlinearity \cite{Hennig,Zhang,Meghoufel,Bulashenko, Ning,Monsivais,K, Ouasti,MGrabowski,Senouci}. Many interesting phenomena are found, such as the localization or
superlocalization properties \cite{K,Ouasti,Lochner,Senouci,Anderson,Peter}, resonant tunneling of electron wave \cite{Monsivais,Senouci,R,Bonilla,Djelti}, multistability in the current-voltage characteristics \cite{Ning} and chaotic behavior caused by nonlinearity \cite{Zhang,Bulashenko,He Bing, Hessari,Wan,NSun}. Transmission
coefficient is related to electronic conductance or resistance \cite{Landauer, R.Landauer,Song}, which plays an important role in the research of electronic
transport properties. For a superlattice system modeled by a nonlinear Kronig-Penney (KP) equation with a homogeneous electric field (linear potential) \cite{Ning,Monsivais,K,Ouasti}, the previous solutions contained some complicated nonlinear maps, and the previous expressions of
transmission coefficient implied the ladder approximation of linear potential or the plane-wave approximation of the Hankel functions. \emph{Here our goal is to establish a concise and exact strategy for studying the electronic distribution and transmission of the system, and to find some novel results}.

The KP model is an analytically solvable
model of a one-dimensional (1D) crystalline solid in which the
electron-nuclei interactions are replaced by contact potentials
of the Dirac-$\delta$ form \cite{Kronig,A}. Such a model is greatly appreciated for its
simplified structure in introducing external fields, which has been shown
to be powerful in studying transport property of particles in the optical \cite{A,B}, graphene \cite{Barbier,AJLee} and semiconductor superlattices \cite{Lochner,MLuo,Monozon,SunNG,Soukoulis1,Szmulowicz}. The linear KP systems can be regarded as the reductions of the corresponding nonlinear KP systems with nonlinearity vanishing. The nonlinear KP
model \cite{Hennig,Ning, K, Monsivais,Ouasti, MGrabowski,Senouci,Wan,NSun,Friedman,DHennig} has wide applications which are partly implied in equivalent treatments of various nonlinear systems. One of the interesting examples is a cold many-atom system with spatially periodic interaction strengths \cite{J, Sakaguchi}, if the periodic functions fit the approximation: ``the KP potential serves
as a good model even for experiments with a single Fourier
component" \cite{B}. Another important example is a kind of discrete nonlinear systems originating from the tight-binding
forms of nonlinear Schr\"odinger equations \cite{Wan,NSun, Morandotti,Montina, Chong, Delyon, Soukoulis,PHawrylak}, which arises in many fields
of physics and can be treated as equivalent systems of the nonlinear KP
model \cite{MGrabowski,Wan,NSun,Soukoulis1}. However, in the presence of the constant field and nonlinearity, exact transmission spectrum of the model is open to question \cite{Ning,Monsivais,K,Ouasti}. Therefore, \emph{our analytical method can be applied to accurately treat transmission problem of many different physical systems}.

In this paper, we apply an integral equation established in Ref. \cite{Hai} to seek concise exact solution of a 1D nonlinear KP model describing the underlying transmission problem of an
electron wave in the doped semiconductor superlattice materials and interacting with a homogeneous electric field \cite{Ning}, which is mathematically similar to that of a nonlinear electrified chain \cite{Monsivais,K,Ouasti}. By using the novel exact solution, we find a new simple nonlinear map with
recurrence relation connecting the transmission coefficient with
boundary conditions and system parameters. According to the recurrence relation, we suggest a scheme for manipulating electronic distribution and transmission
through the adjustments of the system parameters. An interesting phase coherence effect of quantum state is revealed in the strict expression of transmission coefficient which differs from the previous approximate results.
The aperiodic probability distributions, energy spectrum, constant current densities and arbitrary transmission coefficients which include the
approximate zero transmission and total transmission and the multiple transmissions, are illustrated. The method
based on the exact solution render the control strategies more
transparent, which can be extended directly to some nonlinear cold atomic
systems and reduced linear KP superlattices. As an equivalent treatment the suggested control protocol
could also be applied to investigate electron transport in many different
discrete nonlinear systems.

\section{Exact solution of the nonlinear Kronig-Penner model}

We would like to study transmission problem of
an electron wave in doped semiconductor superlattice and interacting with a homogeneous electric field. Due to the nonlinearity of
a self-consistent potential used to describe the interaction of the effective electrons
with charge accumulation in the ultrathin doped layers \cite{Ning}, quantum dynamics of the system is governed by the 1D nonlinear KP model \cite{Ning,Monsivais,K,Ouasti}
\begin{eqnarray}\label{dy2}
&&\Big[-\frac 1 2 \frac{d^{2}}{dx^{2}}-Fx+\sum_{j=1}^L (\beta+\alpha|\psi(x)|^{2})\delta(x-j)\Big]\psi(x)\nonumber \\
&&=E\psi(x).
\end{eqnarray}
Here the spatial coordinate $x$ and probability density $|\psi(x)|^2$ have been normalized in units of the lattice spacing $\lambda$ and its inverse $1/\lambda$. Consequently, the eigenenergy $E$, electric field $F$, $\delta$-lattice potential strength $\beta$ and the nonlinearity intensity $\alpha$ are, respectively, in units of $E_{r}$, $E_{r}/\lambda$, $E_{r}$ and $\lambda E_{r}$ with $E_{r}=\frac{\hbar^{2}}{m\lambda^2}$ being the recoil energy; the integer $L$ is the number of doped layers in the superlattice. For the considered semiconductor superlattice material with lattice spacing $\lambda=2$nm \cite{Ning,K} and electronic effective mass \cite{Shi} $m=0.067 m_e$ with $m_e$ being the electronic rest mass, the recoil energy is $E_r\approx 0.2848$eV. When the nonlinearity intensity is set as zero, Eq. (1) is reduced to a linear KP system \cite{Monozon,SunNG,Soukoulis1,MLuo}. The multistability in the current-voltage characteristics, localization or superlocalization properties and resonant tunneling of electron wave for system (1)
have been studied, by establishing some complicated nonlinear maps and using approximate treatments of the transmission coefficient \cite{Ning,Monsivais,K,Ouasti}.
Here we will seek concise exact solution of the system, and employ them and the strict definition of transmission coefficient to transparently control the electronic distribution and transmission.

Setting $\zeta=\sqrt[3]{2F}(\frac{E}{F}+x)$, obviously, Eq. (1) can
be turned into
\begin{eqnarray}\label{dy3}
\Big(\frac{d^{2}}{d\zeta^{2}}+\zeta\Big)\psi(x)=2\sum_{j=1}^L(\beta+\alpha|\psi(x)|^{2})\delta(x-j)]\psi(x).
\end{eqnarray}
Further we define $z(x)=\frac{2}{3}\zeta^{\frac{3}{2}}=
\frac{2}{3}\sqrt{2F(\frac{E}{F}+x)^3}$, then for $\delta(x-j)=0$ with $x\neq j$, Eq. (2) becomes the Bessel
equation of order $\frac{1}{3}$ with two linearly independent
solutions \cite{Ning}
\begin{eqnarray}\label{dy5}
\varphi(x)&=&\sqrt[3]{z(x)} H^{(1)}_{1/3}[z(x)],\nonumber\\
\phi(x)&=&\frac{i\pi}{4\sqrt[3]{3F}}\sqrt[3]{z(x)} H^{(2)}_{1/3}[z(x)],
\end{eqnarray}
where $H^{(1,2)}_{1/3}[z(x)]$ are the Hankel functions of the first and second
kind. Selection of the constant factor $\frac{i\pi}{4\sqrt[3]{3F}}$ is due to considering the simplicity of the integral equation (4) and its exact solution.

Now we can use the method of integral equation presented in Ref. \cite{Hai, Huang} to construct exact solution
of Eq. (2). Assuming the electric field is applied in the spatial range $0\le x\le L$ of the doped semiconductor, in terms of the functions in Eq. (3), the integral equation corresponding Eq. (2) reads
\begin{eqnarray}\label{dy7}
&&\psi(x)=A_{1}\varphi(x)+B_{1}\phi(x)\nonumber\\
&+&\phi(x)\int_0^x 2\varphi(s)\sum_{j=1}^{[x]}[\beta+\alpha|\psi(s)|^{2}]\delta(s-j)\psi(s)ds\nonumber\\
&-&\varphi(x)\int_0^x 2\phi(s)\sum_{j=1}^{[x]}[\beta+\alpha|\psi(s)|^{2}]\delta(s-j)\psi(s)ds,\ \ \
\end{eqnarray}
where the summations vanish for $x<1$, function $A_{1}\varphi(x)+B_{1}\phi(x)$ is the general
solution of Eq. (2) at $x\ne j$, and sign $[x]$ denotes an integer
obeying $[x]\ge 1$ and $x-1< [x]\le x$. Directly taking second derivative of Eq. (4) and making use of the Wronski determinant
$\varphi\cdot\partial_{x}\phi-\phi\cdot\partial_{x}\varphi=1$, we can easily
prove that the integral equation (4) is completely equivalent to the differential equation (2). Taking the first derivative $\psi_x(x)$ from Eq. (4) and
completing the integrals in $\psi(x)$ and $\psi_x(x)$, \emph{we arrive at the exact solution and its first derivative}
\begin{eqnarray}\label{dy8}
&& \psi(x)=A_{1}\varphi(x)+B_{1}\phi(x)\nonumber\\
&&+2\phi(x)\sum_{j=1}^{n-1}\varphi(j)[\beta+\alpha|\psi(j)|^{2}]\psi(j)\nonumber\\
&&-2\varphi(x)\sum_{j=1}^{n-1}\phi(j)[\beta+\alpha|\psi(j)|^{2}]\psi(j),\nonumber
\end{eqnarray}\label{dy9}
\begin{eqnarray}\label{dy9}
&& \psi_x(x)=A_{1}\varphi_x(x)+B_{1}\phi_x(x)\nonumber\\
&&+2\phi_x(x)\sum_{j=1}^{n-1}\varphi(j)[\beta+\alpha|\psi(j)|^{2}]\psi(j)\nonumber\\
&&-2\varphi_x(x)\sum_{j=1}^{n-1}\phi(j)[\beta+\alpha|\psi(j)|^{2}]\psi(j), \   0 \le x < n\ \ \ \
\end{eqnarray}
with $n-1 =[x]$. Note that the summations vanish for $x<1$. Such exact expressions can be rewritten in the forms
\begin{eqnarray}\label{dy6}
\psi(x)&=&A_{n}\varphi(x)+B_{n}\phi(x),\nonumber\\
\psi_x(x)&=&A_{n}\varphi_x(x)+B_{n}\phi_x(x),   \   0 \le x < n,
\end{eqnarray}
where the integral constants $A_{n}$ and $B_{n}$ are related to the electronic probability distribution and transmission coefficient, which satisfy the relation between $(A_1,B_1)$ and $(A_n,B_n)$
\begin{eqnarray}\label{dy7}
A_{n}= A_{1}-2\sum_{j=1}^{n-1}\phi(j)[\beta+\alpha|\psi(j)|^{2}]\psi(j),\nonumber\\
B_{n}= B_{1}+2\sum_{j=1}^{n-1}\varphi(j)[\beta+\alpha|\psi(j)|^{2}]\psi(j).
\end{eqnarray}
This relation implies the nonlinear map connecting $(A_{n+1},B_{n+1})$ with $(A_n,B_n)$,
\begin{eqnarray}\label{dy8}
A_{n+1}&=& A_{n}-2\phi(n)[\beta+\alpha|\psi(n)|^{2}]\psi(n),\nonumber\\
B_{n+1}&=& B_{n}+2\varphi(n)[\beta+\alpha|\psi(n)|^{2}]\psi(n)
\end{eqnarray}
for $2\le n+1\le L$. Here $\psi(n)=A_{n}\varphi(n)+B_{n}\phi(n)$ is related to $A_n$ and $B_n$ by Eqs. (6) and (3). \emph{The recurrence relation of Eq. (8) is very simple compared to the previously established nonlinear or linear maps} without the ladder approximation \cite{Ning,SunNG} or with the ladder approximation \cite{Monsivais,K,Ouasti,Soukoulis1,Soukoulis}, because of the simplicity of our exact solution.
Given Eq. (8), one can easily prove that the exact solution satisfies the continuity condition of $\psi(x)$ and the jumping condition of $\psi_x(x)$ at any spatial singular point $x=n$ of Eq. (2), \cite{Ning,SunNG}
\begin{eqnarray}\label{dy9}
&& \psi(n^+)-\psi(n^-)\nonumber\\
&&=A_{n+1}\varphi(n)+B_{n+1}\phi(n)-[A_{n}\varphi(n)+B_{n}\phi(n)]=0; \nonumber\\
&& \psi_x(n^+)-\psi_x(n^-)\nonumber\\
&& =A_{n+1}\varphi_x(n)+B_{n+1}\phi_x(n)-[A_{n}\varphi_x(n)+B_{n}\phi_x(n)] \nonumber\\
&& =2[\beta+\alpha|\psi(n)|^{2}]\psi(n)[\phi_x(n)\varphi(n)-\varphi_x(n)\phi(n)] \nonumber\\
&& =2[\beta+\alpha|\psi(n)|^{2}]\psi(n),
\end{eqnarray}
where $n^+$ and $n^-$ denote, respectively, $n+\epsilon$ and $n-\epsilon$ for $0< \epsilon \ll 1$. In the calculations, the continuity of $\phi(x),\ \varphi(x)$ and Wronski determinant $\phi_x(n)\varphi(n)-\varphi_x(n)\phi(n)=1$ have been employed. The agreement between Eq. (9) and the direct first integration of Eq. (1) over a delta confirms the correctness of the exact solution.

From Eqs. (3), (6) and (8) we find that for the given boundary conditions $\psi (0)=A_1\varphi(0)+B_1\phi(0),\ \psi (L)=A_L\varphi(L)+B_L\phi(L)$ and a set of fixed system parameters, constants $A_n, B_n$ and the corresponding eigenenergy $E$ can be obtained for any $n$ obeying $2\le n+1\le L$. Thus Eq. (8) contains the connection of probability distribution and transmission coefficient with
boundary conditions and system parameters. Combining the analytical expressions with simple numerical calculations, we suggest
a scheme for manipulating electronic distribution and transmission
by adjusting system parameters as follows.

\section{Manipulating electronic distribution and transmission}

Considering an incident electron wave coming from the left towards
the nonlinear doped semiconductor superlattice extending over $L$ lattice sites and noticing the electric field range $0\le x\le L$, our transmission problem can be defined as the following \cite{Delyon,Soukoulis}
\begin{eqnarray}\label{dy10}
\psi_l(x)&=&R_{0}e^{ik(x+a)}+R_{1}e^{-ik(x+a)} \ \ \text{for}\ \ x\leq 0,\nonumber\\
\psi_c(x)&=&A_{n}\varphi(x)+B_{n}\phi(x) \ \ \text{for}\ \ 0\le x\leq L,\nonumber\\
\psi_r(x)&=&Te^{ik(x+b)} \ \ \text{for}\ \ x\ge L.
\end{eqnarray}
Here the left wave function $\psi_l(x)$ is a superposition of the incident plane wave $R_{0}e^{ik(x+a)}$ and reflected plane wave $R_{1}e^{-ik(x+a)}$ with the corresponding real amplitudes $R_{0}$, $R_{1}$, wave vector $k$ and phase $a$. The right plane wave $\psi_r(x)$ is the transmitted wave with the real amplitude $T$, wave vector $k$ and phase $b$. The amplitudes, wave vector and phases are normalized in units of $1/\sqrt{\lambda},\ 1/\lambda$ and $\lambda$, respectively. In the range $0\le x\leq L$ of the doped semiconductor sample, the center electronic state $\psi_c(x)$ obeys Eqs. (6) and (8). The signs ``$\le$" and $``\ge"$ in Eq. (10) describe the continuity of wavefunction at the sample boundaries $x=0,L$.

It is well-known that for given $\psi_l(x),\ \psi_r(x)$ and fixed system parameters $(F, \alpha, \beta, L)$, a set of electronic states and eigenenergies \cite{Ning,Wan} can be determined by the boundary conditions of the sample at $x=0,L$. The usual treatment of a transmission problem consists of finding the reflected and transmitted
amplitudes, in terms of the incident amplitude and energy. The approximate transmission coefficient and current-field characteristic of the system have been investigated  based on some given parameters and fixed energies \cite{Ning,K}. The method to invert the problem by fixing the output and then calculating the input has also been employed \cite{Monsivais,Ouasti,Delyon,Wan,PHawrylak}. We are interested in the electronic exact distribution and transmission by solving the \emph{inverse problem}: for priori prescribed incident wave amplitude $R_0$ and reflected wave amplitude $R_1$ with an adjustable phase $a$ and for a set of fixed values of eigenenergy and system parameters $(E,\ F, \alpha, \beta, L)$, we seek the electronic states $\psi_c(x)$, $\psi_r(x)$ and suitable superlattice length $L$ which fit the boundary conditions  $\psi_c(0)=\psi_l(0)$ and $\psi_c(L)=\psi_r(L)$. Noticing the conservation formula of probability \cite{Delyon,Soukoulis} $T^2=R_{0}^2-R_{1}^2$, clearly, the results based on the inverse problem can display useful relations connecting the electronic distribution and transmission coefficient with the system parameters.

In fact, from the exact solution (6) with Eqs. (3) and (8) we know that for given parameters $(E, F, \alpha, \beta)$ and determined constants $A_1, B_1$, one can obtain the wavefunction $\psi_c(x)$ of the continuously varying $x$ or discrete coordinate $x=n=1,...,L$. For the priori prescribed incident and reflected waves, the constants $A_1, B_1$ are determined by the left-boundary condition of Eq. (10) at $x=0$ and the transmission amplitude $T$ is adjusted by the probability conservation. For some $L$ values satisfying the right-boundary condition $\psi_c(L)= \psi_r(L)$, the obtained $\psi_c(x)$ and $\psi_r(x)$ are the correct solutions of the inverse problem. Through such a method we will reveal that the exact solution enables us to conveniently manipulate the electronic distribution $|\psi_c(x)|^2$ and the well-defined transmission coefficient \cite{Soukoulis,NSun} $t=T^2/R_0^2$.

According to the above analysis, after fixing a set of parameters $(E,\ F, \alpha, \beta)$ and priori prescribing plane wave $\psi_l(x)$ with known parameters $(R_0, R_1, k)$ and $\sqrt{R_{0}^2-R_{1}^2}=T$, we solve the transmission problem along the two steps:

\emph{Step 1}. We firstly determine the constants $A_1, B_1$ by adjusting phase $a$ and solving the left-boundary equation of Eq. (10) at $x=0$,
\begin{eqnarray}\label{dy13}
R_0 e^{ika}+R_1 e^{-ika}=A_{1}\varphi(0)+B_{1}\phi(0),
\end{eqnarray}
where functions $\varphi(x)$ and $\phi(x)$ are fixed by Eq. (3). To simplify, we select $R_0, R_1, k, a$ as positive real constants, and real $A_1$ and imaginary $B_1$ to obey $A_1=|A_1|,\ B_1=-i|B_1|$. Thus Eq. (3) gives the relationship of arguments
\begin{eqnarray}\label{dy12}
\arg(A_{1}\varphi)=\arg(\varphi)=-\arg(B_{1}\phi)=-\arg(-i\phi)
\end{eqnarray}
at any $x$ point. Note that for different values of parameters $E,\ F$, Eq. (3) may give positive or negative argument $\arg[\varphi(0)]$. \emph{In the case of $\arg[\varphi(0)]>0$}, Eqs. (11) and (12) mean that
\begin{eqnarray}\label{dy14}
&&\arg[\varphi(0)]=k a=-\arg[-i\phi(0)],\nonumber\\
&&|A_1\varphi(0)|=R_0,\ \ \ |B_1\phi(0)|=R_1.
\end{eqnarray}
The first of Eq. (13) gives the undetermined constant $a$, and the other two equations result in values of $|A_1|$ and $|B_1|$.
\emph{In the case of $\arg[\varphi(0)]<0$}, the formulas similar to Eq. (13) are produced as
\begin{eqnarray}\label{dy14}
&&\arg[\varphi(0)]=-k a=-\arg[-i \phi(0)],\nonumber\\
&&|A_1\varphi(0)|=R_1,\ \ \ |B_1 \phi(0)|=R_0,
\end{eqnarray}
which give the different constants $a$, $|A_1|$ and $|B_1|$. Given $A_1=|A_1|, B_1=-i|B_1|$, from the exact solution (6) with Eqs. (3) and (8) we derive the wavefunction $\psi_c(x)$ of the discrete coordinate $x=j=1,...,L$.

\begin{figure*}[htp]\center
\includegraphics[height=1.6in,width=3.0in]{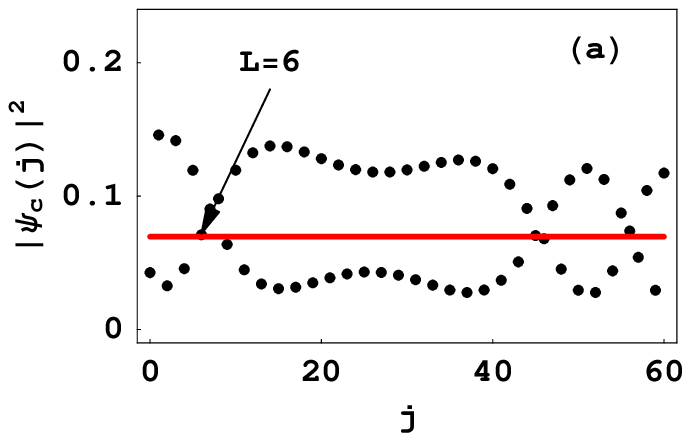}
\includegraphics[height=1.6in,width=3.0in]{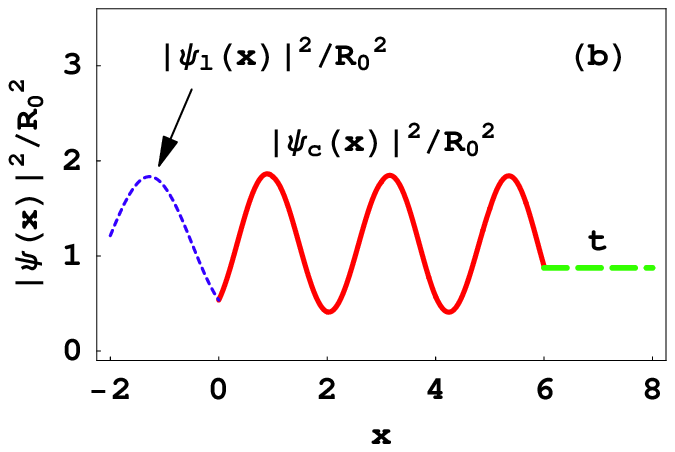}
\caption{\scriptsize{(Color online)  Plots showing the right-boundary condition $\psi_c(L)=
\psi_r(L)$ in (a) and the electronic distribution for the case of general
transmission in (b). The parameters are chosen as
$E=1$, $F=0.01$, $\beta=0.025$, $R_0=0.2822$, $k=1$, $a=1.2766$, $\alpha=0.015$ and $R_1=0.1$. The dotted curve in (a) describes the probability density $|\psi_c(j)|^{2}$ as a function of the discrete coordinate $j$ and the
solid line indicates the value of the probability density $|\psi_{r}(x)|^2=T^2\approx 0.0696$. Taking the dots coinciding with the solid line as the boundary points $L= 6, 45,
46$, the system satisfies the right-boundary condition $|\psi_c(L)|^{2}=T^{2}$. In (b) with lattice length $L= 6$, the short dashed curve, solid curve and long dashed line describe the relative probability density
$|\psi(x)|^{2}/R_{0}^{2}$ in the three different spatial ranges, which shows the periodic, near-periodic and constant distributions, respectively. Hereafter all the quantities plotted in the figures are
dimensionless.}}
\end{figure*}

\emph{Step 2}. We then determine appropriate values of the length $L$ and phase $b$ which fit the right-boundary condition of Eq. (10) at $x=L$,
\begin{eqnarray}\label{dy16}
\psi_c(L)=A_L \varphi(L)+B_L \phi(L)=Te^{ik(L+b)},
\end{eqnarray}
where the continuity condition of $\psi_c(x)$ at $x=L$ has been adopted. It is interesting to find that different $L$ values can fit the same amplitude equation of Eq. (15),
\begin{eqnarray}\label{dy18}
|A_L \varphi(L)+B_L \phi(L)|=T,
\end{eqnarray}
because of the \emph{phase coherence} between the wave components $A_L \varphi(L)$ and $B_L \phi(L)$. Note that in Eq. (16) \emph{the amplitude $T$ of transmitted wave is an exact result, in contrast to the earlier approximate treatments} \cite{Ning,K, Monsivais,Ouasti}. In Ref. \cite{Ning}, only the $A_L$ term is counted due to the plane-wave approximation of the Hankel functions, so such a phase coherence is ignored. For a $L$ value derived from Eq. (16), the corresponding value of phase $b$ is determined easily by the argument equation of Eq. (15)
\begin{eqnarray}\label{dy17}
\arg[A_L \varphi(L)+B_L \phi(L)]=k(L+b).
\end{eqnarray}
Here different length $L$ is associated with different phases $\arg[\psi_c(L)]$ and $b$, since a single $T$ value in Eq. (16) can be related to multiple transmitted waves of different phases in Eq. (17). So far the electron waves $\psi_c(x)$ and $\psi_r(x)$ have been determined completely for any obtained $L$ value. The corresponding probability density and current density can be calculated. Such a length $L$ may be used in preparation of the doped semiconductor superlattice.

It is worth noting that in the inverse problem the transmission coefficient $t=T^2/R_0^2=1-R_1^2/R_0^2$ has been priori prescribed \cite{Monsivais} by using the amplitudes $R_0, R_1$ of the incident and reflected waves. Therefore, we can take three different cases of the same incident wave as examples to show the effects of the system parameters on the electronic distribution and transmission.

\emph{Case 1: General transmission with $0< t <1$}. At first, we arbitrarily
take the eigenenergy and system parameters $(E, F, \alpha,
\beta)=(1, 0.01, 0.015, 0.025)$ and the left-plane-wave parameters
$(R_0, R_1, k)=(0.2822, 0.1, 1)$ to prescribe the transmission amplitude
$T= \sqrt{R_{0}^2-R_{1}^2}=0.2639$ and transmission coefficient
$t=1-R_1^2/R_0^2\approx0.8744$. Then the constants $A_1=0.2674,
B_1=-0.2985i, a= 1.2766$ are derived from the above step 1, and the
appropriate values of the length are given by Eq. (16) as $L= 6, 45,
46,...$, as shown in Fig. 1(a). From Eq. (17) the corresponding phases are obtained as $b=
-6.5656, -44.5328, -46.5050, ... $. In Fig. 1(a) we plot the probability density $|\psi_c(j)|^2$ of the discrete coordinate $x=j=1,...,L$ as the dotted curve, where
the solid line indicates the value $T^2\approx 0.0696$. Clearly, there exist some dots coinciding with the solid line, which mean that the corresponding multiple values of length $L$ satisfy the right-boundary condition $|\psi_c(L)|^{2}=T^{2}$. The used energy $E=1 (E_r)\approx 0.2848$eV is in the same order as those adopted in Ref. \cite{Ning,K}. For the obtained
lattice length $L=6$ and the above other parameters, we plot the exact
electronic distribution in Fig. 1(b) by using Eq. (10) and the exact
solution (6) with Eqs. (3) and (8). Here the relative probability densities $|\psi_l(x)|^2/R_0^2$ and $|\psi_c(x)|^2/R_0^2$
are plotted by the short dashed line and solid line, respectively, and the transmission coefficient $t$
is labeled by the long dashed line. In the spatial range $x\le 0$, the superposition of the incident wave and reflected wave with wave vector $k$ results in the spatially periodic density $|\psi_l(x)|^2$. While the probability density of the transmitted plane wave is a constant at the right side with $x\ge L=6$. Inside the doped semiconductor superlattice, $0\le x \le L$, the exact electronic distribution $|\psi_c(x)|^2$ is near-periodic, because of the small difference $F\Delta x=FL=0.06$ of the aperiodic linear potential. At the boundary points $x=0, L$, continuity of the wavefunction is displayed in Fig. 1(b). The \emph{phase coherence} between two wave components of $\psi_c(L)$ in Eq. (15) leads to the suitable transmission coefficient.
Inside the doped material, the probability current density reads \cite{MGrabowski,MLuo} $\overrightarrow{j}
=0.5i[\psi_c(x)\nabla\psi_c^{*}(x)-\psi_c^{*}(x)\nabla\psi_c(x)]\frac{\hbar}{m\lambda^2}
=-0.098 \omega_r \overrightarrow{e_x}$ with $\omega_r=E_r/\hbar$ being the recoil frequency and
$\overrightarrow{e_x}$ the unit vector in $x$ direction. The
negative current density means the electronic motion along the $x$
direction.

\begin{figure*}[htp]\center
\includegraphics[height=1.6in,width=3.0in]{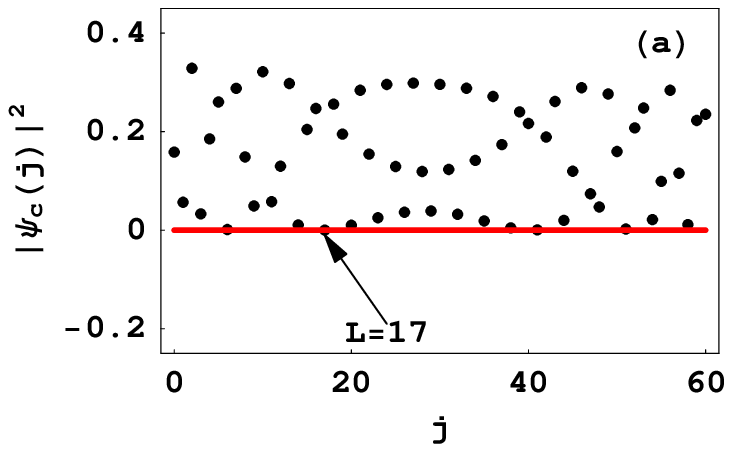}
\includegraphics[height=1.6in,width=3.0in]{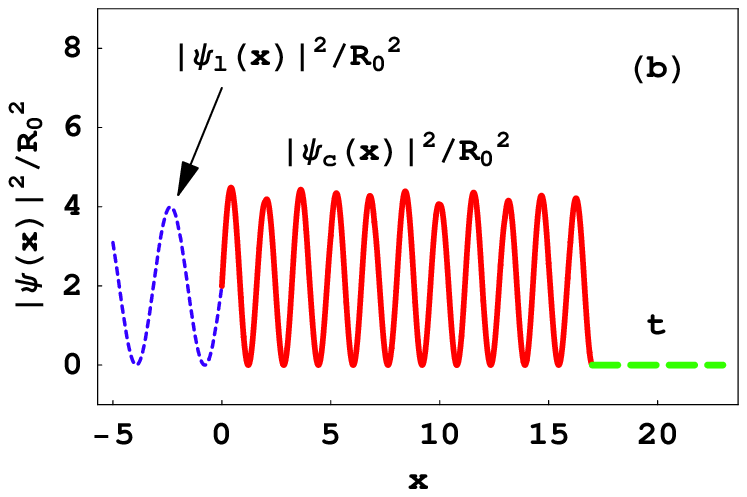}
\caption{\scriptsize{(Color online)  Plots showing the right-boundary condition and the electronic distribution in the zero transmission case for the parameters $E=2$,
$F=0.011$, $\alpha=0.1$, $\beta=0.1$, $R_0=0.2822$, $k=1$, $a=1.3539$ and $R_1=0.2821$.
(a) probability density $|\psi_c(j)|^{2}$ versus $j$, where the solid line
labels the value of $T^{2}<10^{-4}$ and the lattice lengths obeying the right-boundary
condition read $L= 6, 17, 38, 41, 51$. (b) The relative probability density versus $x$ for the obtained
lattice length $L= 17$, which shows the periodic, near-periodic and zero distributions in the three different spatial ranges, respectively. }}
\end{figure*}

\emph{Case 2:  Approximate zero transmission with t $\approx 0$}. Similarly we
take the eigenenergy and system parameters as $(E, F, \alpha,
\beta)=(2, 0.011, 0.1, 0.1)$ and the left-plane-wave parameters
$(R_0, R_1, k)=(0.2822, 0.2821, 1)$ to yield the transmission
amplitude $T\approx0.0075$ and
transmission coefficient $t=0.00071 \approx 0$. From the above step 1, the boundary
constants are derived as $A_1=\sqrt{0.78}, B_1=-i\sqrt{0.13}, a= 1.3539$, and the appropriate values of the
lattice length are given by step 2 as $L= 6, 17, 38, 41, 51,...$, as
shown in Fig. 2(a); and the corresponding phases values read $b= -5.9973,\
-20.1340,\ -41.1402$, $-41.0037,\ -50.9981,... $.
With the similar approach used in Fig. 1($b$), for the obtained lattice length $L= 17$ and the other parameters of case 2, we plot the exact
electronic distribution in Fig. 2(b), where the relative probability densities $|\psi_l(x)|^2/R_0^2$, $|\psi_c(x)|^2/R_0^2$ and transmission coefficient $t$ are periodic, near-periodic and constant, respectively.  The \emph{coherence destruction} between the wave components of $\psi_c(L)$ results in the approximate zero transmission.
In the zero transmission case, we get the probability current
density $j \approx 0$ of electron wave, which means that the incident wave is completely reflected and electron transport cannot occur in
the doped material.

\begin{figure*}[htp]\center
\includegraphics[height=1.6in,width=3.0in]{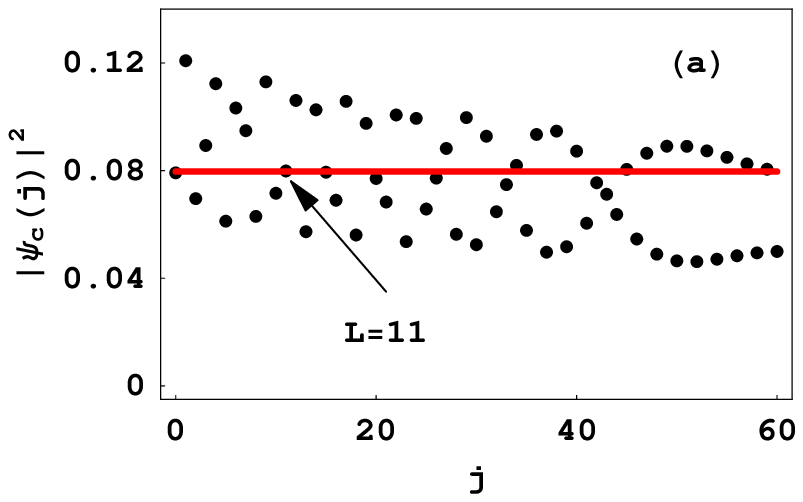}
\includegraphics[height=1.6in,width=3.0in]{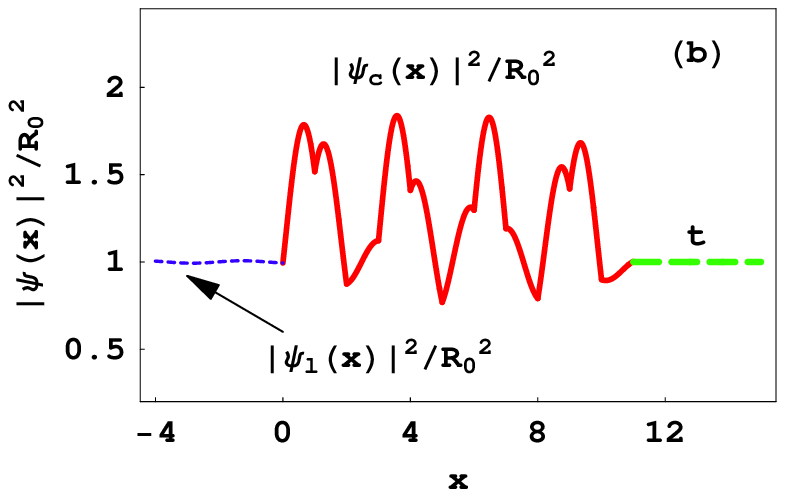}
\caption{\scriptsize{(Color online)  Plots showing the right-boundary condition and the electronic distribution in the total
transmission case for the parameters $F=0.01$, $k=1$, $\alpha=0.02$,
$\beta=0.415$, $R_0=0.2822$, $R_1=0.001$, $E=1$, $a=1.2766$. In plot of $|\psi_c(j)|^{2}$ versus $j$ of Fig. 3(a), the
solid line at $T^{2}=0.0796$ indicates the different lengths satisfying
the right-boundary condition $|\psi_c(L)|^{2}=T^{2}$ as $L= 11,
15, 45, 59$. In Fig. 3(b), the relative probability density versus $x$ is plotted for the obtained lattice length $L= 11$, where the aperiodic electronic distribution in the doped material and the approximately same incident and transmitted plane wave are displayed. }}
\end{figure*}

\emph{Case 3:  Approximate total transmission with $t \approx 1$}. When the eigenenergy and system parameters $(E, F, \alpha,
\beta)=(1, 0.01, 0.02, 0.415)$ and the left-plane-wave parameters
$(R_0, R_1, k)=(0.2822, 0.001, 1)$ are adopted, the transmission
amplitude $T\approx0.2822$ and
transmission coefficient $t\approx1$ are prescribed. Then the
constants $A_1=0.0027, B_1=-0.2985i, a= 1.2766$ are given by the
above step 1. The above step 2 leads to the appropriate values of the lattice length as $L= 11, 15, 45, 59,...$ which are shown in Fig. 3(a).
The corresponding phase values become $b= -9.8816,\ -12.1483,\ -45.0147$, $-61.0805,... $ based on Eq. (17). For the obtained lattice
length $L= 11$ and the above other parameters, we also plot the exact
electronic distribution in Fig. 3(b). Here the relative probability densities of both the incident and transmitted waves are approximately one, and aperiodicity of the electronic distribution is intuitive inside the doped material. The approximate total transmission is induced by the \emph{coherence construction} between the wave components of $\psi_c(L)$. Similarly, we calculate the probability current
density $j \approx -0.113 \omega_r$ in the case of total transmission, which
means the electron transport occurs along the $x$ direction in the doped material.

Comparing the probability current densities in the three cases with the same incident wave and different transmission coefficients, we find that the current densities in the superlattice material are positively related to the transmission coefficients. While the transmission coefficients are associate with the conductance $G$ of the doped material, through the Landauer formula \cite{Anderson,Ning,R.Landauer} $G \sim t/(1-t)$. The approximate zero transmission and total transmission correspond to the approximate zero conductance and zero resistance, respectively. Generally, the transmission coefficients are adjusted by the system parameters $(F, \alpha,
\beta, L)$, which means that the conductance of the doped material can be controlled by the external fields. The exact solution (6) with Eqs. (3) and (8) supplies a more
transparent control strategy. The dependence of transmission coefficient on material length is one of interesting topics \cite{K,Ouasti,PHawrylak}. As an instance, we investigate the effects of material length on transmission coefficient and eigenenergy as follows.

\begin{figure*}[htp]\center
\includegraphics[height=1.6in,width=3.0in]{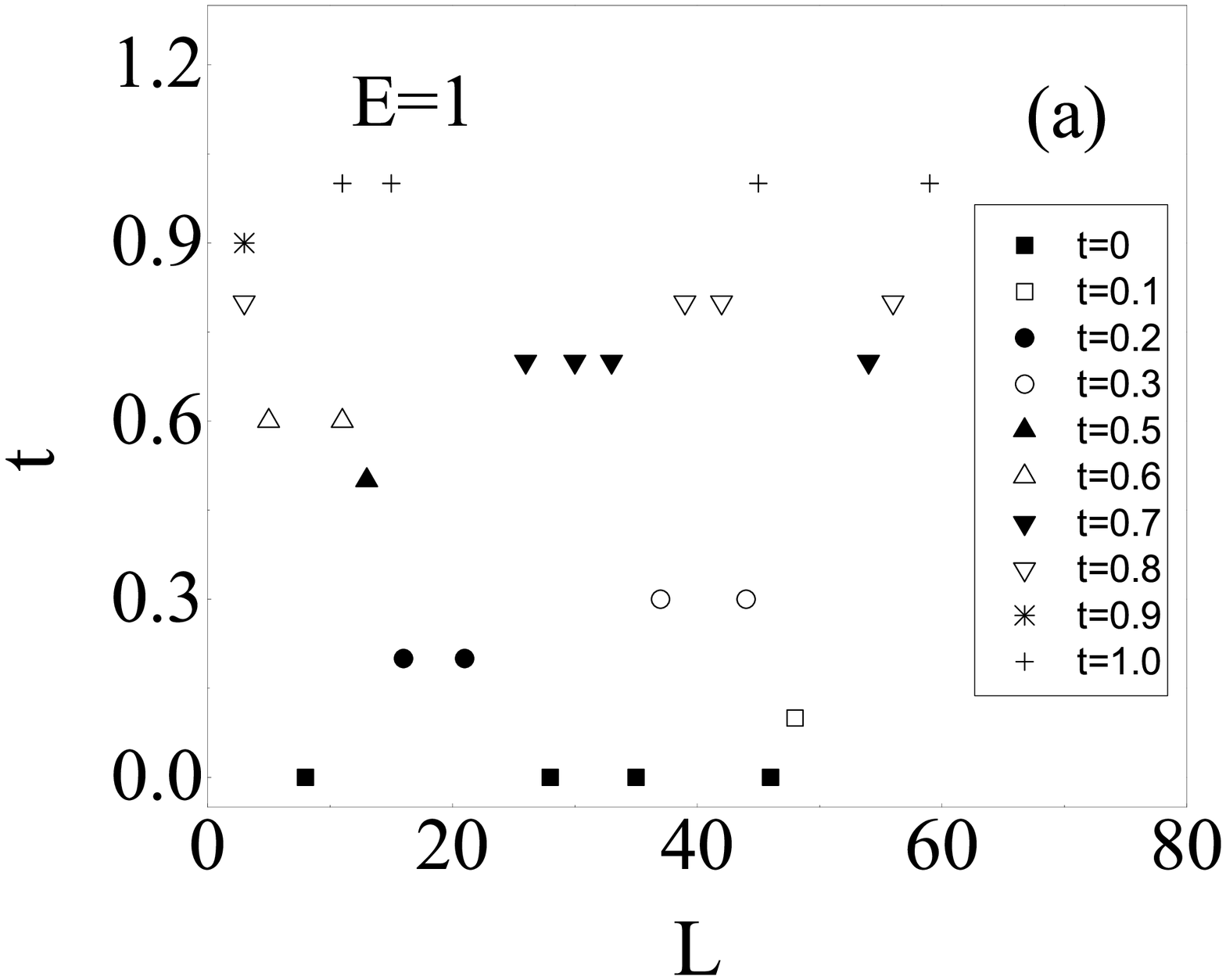}
\includegraphics[height=1.6in,width=3.0in]{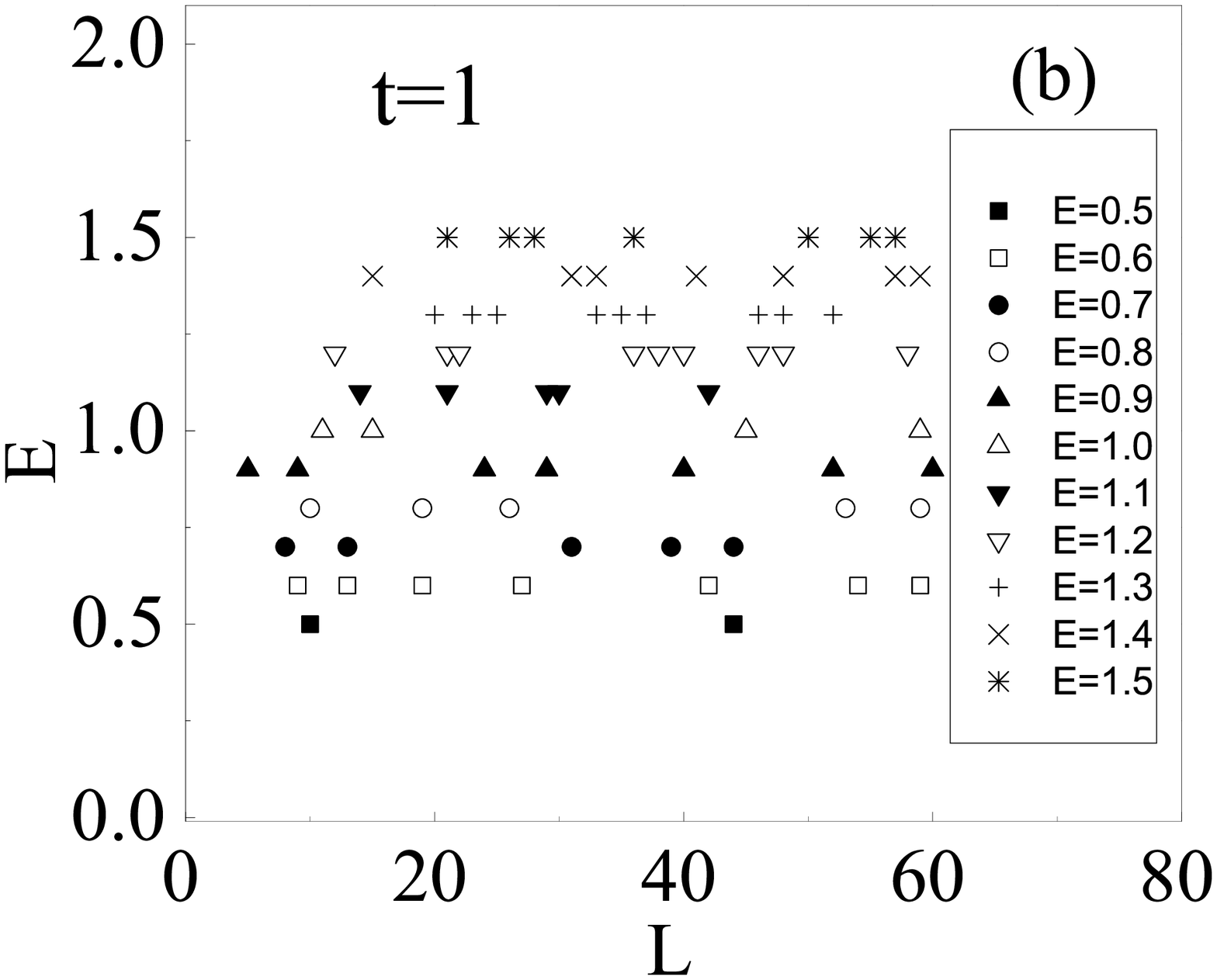}
\caption{\scriptsize{ (a) Transmission coefficient $t$ versus $L$ for eigenenergy $E=1$; (b) Eigenenergy $E$ versus $L$ for $t\approx 1$ and $R_1=0.001$. The other parameters are chosen as $F=0.01$, $k=1$, $\alpha=0.02$, $\beta=0.415$, $R_0=0.2822$. }}
\end{figure*}

\emph{Transmission spectrum and multiple transmissions}. In order to show the connections of transmission coefficient $t$ and eigenenergy $E$ with the material length $L$, we plot Fig. 4 for the system parameters $F=0.01$, $\alpha=0.02$, $\beta=0.415$ and the same incident wave with amplitude $R_0=0.2822$ and wave vector $k=1$. In Fig. 4(a) with $E=1$, we show $t-L$ relation as
\begin{eqnarray}
[t,\ (L)]&=&[0, (8, 28, 35, 46)],\ [0.1, (48)],\ [0.2, (16, 21)], \nonumber
\\ && [0.3, (37, 44)],\ [0.5, (13)],\ [0.6, (5, 11)], \nonumber
\\ && [0.7, (26, 30, 33, 54)],\ [0.8,(3, 39, 42, 56)], \nonumber
\\ && [0.9,(3)],\ [1, (11, 15, 45, 59)] \nonumber
\end{eqnarray}
for $L\le 60$. Clearly, in a certain region of $L$, multiple $L$ values correspond to a single $t$ value \cite{K}, while every $L$ value is associated with different phases $\arg[\psi_c(L)]$ and $b$ of quantum states by Eq. (17). Such an quantum phase effect is fundamentally important in transmission problem of a quantum system. We then observe that one of some $L$ values corresponds to multiple possible $t$ values, such as
\begin{eqnarray}
[L,\ (t)]=[3, (0.8, 0.9)],\ [11, (0.6, 1)]. \nonumber
\end{eqnarray}
This multivaluedness of $t$ means that the multiple transmissions may occur in the exact treatment, due to that the linear superposition principle is no longer valid in the nonlinear case, so $T$ is not uniquely defined by $R_0$ \cite{Ning, Monsivais}. Note that the chosen nonlinearity intensity $\alpha=0.02 (\lambda E_{r})\approx0.1136 {\AA}$ eV is sufficiently weak to lose sight of the multiple transmissions in the approximate treatment without the considered phase coherence \cite{Ning}.

\emph{Energy spectrum}. In Fig. 4(b), we exhibit $E-L$ relation for the total transmission with $t=1$ and $L\le 60$ as
\begin{widetext}
\begin{eqnarray}
[E,\ (L)]&=&[0.5, (10, 44)],\ [0.6, (9, 13, 19, 27, 42, 54, 59)], [0.7, (8, 13, 31, 39, 44)],\ [0.8, (10, 19, 26, 53, 59)], \nonumber
\\ && [0.9, (5, 9, 24, 29, 52, 60)],\ [1, (11, 15, 45, 59)], [1.1, (14, 21, 29, 30, 42)], [1.2, (12, 21, 22, 36, 38, 40, 46, 48, 58)],  \nonumber
\\ && [1.3, (20, 23, 25, 33, 35, 37, 46, 48, 52)],  [1.4, (15, 31, 33, 41, 48, 50, 57, 59)], \nonumber
\\ && [1.5, (21, 26, 28, 36, 50, 55, 57)]. \nonumber
\end{eqnarray}
The correspondence between multiple $L$ values and a singe $E$ value means the similar phase effect of quantum state $\psi_c(L)$. We also find that one of some $L$ values corresponds to multiple possible $E$ values, such as
\begin{eqnarray}
[L,\ (E)]&=&[9, (0.6, 0.9)],\ [10, (0.5, 0.8)],\ [13, (0.6, 0.7)],  [15, (1, 1.4)],\ [19, (0.6, 0.8)],\ [21, (1.1, 1.2, 1.5)],\ [26, (0.8, 1.5)], \nonumber
\\ && [29, (0.9, 1.1)],\ [31, (0.7, 1.4)],\ [33, (1.3, 1.4)], [36, (1.2, 1.5)],\ [42, (0.6, 1.1)],\ [44, (0.5, 0.7)],\ [46, (1.2, 1.3)],  \nonumber
\\ && [48, (1.2, 1.3, 1.4)],\ [50, (1.4, 1.5)], [52, (0.9, 1.3)],\ [57, (1.4, 1.5)],\ [59, (0.6, 0.8, 1, 1.4)]. \nonumber
\end{eqnarray}
\end{widetext}
In the ladder approximation, the similar multivaluedness of energy was associated with the nonlinear Stark ladder resonance which leads to electronic resonant tunneling between minibands \cite{Monsivais}. For $t=1$ and any fixed $L$ our multiple energies are based on the exact solution and can deviate from the energy resonant peaks of the approximate transmission coefficient.

Similarly, we also can derive the relations between transmission coefficient and other system parameters. Given these relations, we can provide different schemes to manipulate electronic distribution and transmission,
through adjustments of the system parameters.

\section{Conclusion}

We have used a nonlinear KP model to study the manipulations of probability distribution and
transmission of an electron wave in a doped semiconductor
superlattice and interacting with a homogeneous electric field. By applying an integral equation to seek concise exact solution of the system, we have established a new simple nonlinear map with
recurrence relation connecting the strictly-defined transmission coefficient with the
boundary conditions and system parameters, in contrast to the earlier complicated recurrence equations and approximate transmission coefficients. Based on our recurrence relation and boundary conditions, we have testified an interesting phase coherence effect of quantum state in the strict expression of transmission coefficient, by which for some different system parameters we have obtained the similar aperiodic electronic
distributions, different energy spectra and constant current densities, and arbitrary transmission coefficients including the
approximate zero transmission and total transmission, and the multiple transmissions.

The method based on the concise exact solution render our control strategies more transparent, which not only can be applied to some nonlinear cold atomic
systems  \cite{J, Sakaguchi} but also can be extended to investigate electron transport in different discrete nonlinear systems \cite{Wan,NSun,Morandotti,Montina, Chong,Delyon, Soukoulis,PHawrylak}. By letting the nonlinearity strength be zero, our analytical method also can be directly applied to a lot of linear KP superlattice systems subjected a dc electric field, including the optical \cite{A,B}, graphene \cite{Barbier,AJLee} and semiconductor superlattice systems \cite{MLuo,Monozon,SunNG,Soukoulis1}.
The considered model can even be modified to study other heterostructures in an electric field, such as a superlattice consisting of alternative n- and p-type doped layers \cite{Ning}, and a nonlinear electrified chain with a disorder potential \cite{K,Ouasti}.

\section*{Acknowledgments}
This work was supported by the NNSF of China under Grant Nos. 11175064, 11204077 and 11475060, and the Construct Program of the National Key Discipline of China.

\end{document}